\documentstyle[12pt]{article}                   
\begin{document}
\renewcommand{\thesection}{\Roman{section}}
\baselineskip 20pt
\input feynman.tex
{\hfill\bf PUTP-96-03}
\vskip 2cm
\begin{center}
\par
{\large\bf  Top Quark Decays into Heavy Quark Mesons}
\vskip 7mm
\centerline{Cong-Feng Qiao~~~ Chong Sheng Li~~~ Kuang-Ta Chao }
{\small\it $ CCAST (World~ Laboratory),~~Beijing~ 100080, ~~P.R.C.$}\\
{\small\it $ Department~ of~ physics,~~Peking~ University ~~Beijing~ 100871,~~ P.R.C.$}
\end{center}
\vskip 0.7in
\begin{center}
\begin{minipage}{120mm}
\begin{center}{\bf Abstract}\end{center}
\par
~~~

\par
For top quark decays into heavy quark mesons $\Upsilon$ and $\bar{B}_c^*$ ,
a complete calculation to the leading order both in QCD coupling constant
$\alpha_s$ and in $v$, the typical velocity of the heavy quarks inside the 
mesons, is performed. Relatons between the top quark mass and the decay
branching ratios are studied. Comparion with the results which are obtained 
by using the quark frangmentation functions is also discussed. The branching 
ratios are consistent (within a factor of $2\sim 3 $) with that obtained using  
fragmentation functions at $m_t\sim 150$ GeV.

\par  
~~
\par
PACS number(s):$ 12.38 Bx$~ $13.60 Le$~ $14.65 Ha$
\end{minipage}
\end{center}
\vskip 7cm
\begin{center}
{Submitted to Phys. Rev.D}
\end{center}
\vfill\eject\pagestyle{plain}\setcounter{page}{1}

\begin{center}\section{Introduction}\end{center}
\par  
~~~
 The success of Standard Model (SM)\cite{s1}\cite{s2}\cite{s3} suggests 
that the top quark must 
exit \cite{s4}. Recently, from the direct search at the Tevatron, the {\bf CDF}
and {\bf D0} groups confirmed the existence of heavy top quark\cite{s5}\cite{s6}, 
which has a mass of $(176 \pm 8 \pm 10) $ GeV or
$(199^{+19}_{-21} \pm 22 )$ GeV. Then 
next the experimental studies will focus on the determination of its properties.
Among them, the precision 
measurement of the top quark mass and of the production cross 
sections and distributions will certainly be in the first studies of interest.
Since the fermion mass generation can be closely related to the electroweak 
symmetry breaking, one expects to find some residual effects of the breaking 
in accordance with the generated mass especially in the much heavier top quark
sector. Therefore, it is important to study the top quark system as a direct 
tool to probe new physics effects\cite{s4}. 

As the new found top quark mass is much heavier than that one previously 
expected, its decay widths in many processes should be reevaluated. With future 
upgrade  and increases in total integrated luminosity, experiments at the
Fermilab Tevatron will carry a carefull investigation of the production and 
decay of top quark. With the operation of CERN Large Hadron Collider (LHC), 
one expects to obtain roughly $ 10^7-10^8~  t{\bar t}$ pairs per year \cite{LHC}
and then one may be able to observe various top decay channels, which will give 
further tests of the SM and heavy meson production mechamisms.
On the other hand, the study of 
heavy quarkonium and heavy meson production in turn provides a ground to 
precisely test the 
perturbative quantum chromodynamics (PQCD). 

Within past few years, much progress has been made in the study of production 
mechanism of heavy quarkonium and heavy mesons with large transverse momentum at 
high energies.  The parton 
fragmentation in the heavy quarkonium and heavy meson
production has been carefully studied \cite{function}\cite{bc2} and  the 
universal fragmentation functions for different spin
and angular momentum states have been obtained. 
With these functions, the branching ratios of 
heavy quarkonium and heavy meson production for various channels may be easily 
obtained if the fragmentation mechanism is in dominance. 
Based on the new factorization formulas\cite{nrqcd} and the production mechanism, 
one may explain some new experiment results\cite{oct}.
Hence, in what  
energy region and in what degree of precision the 
fragmentation mechanism works in
a specific process should be concerned.

The most important concequence of a heavy top quark is that, because its 
lifetime is short, after created in free state it does not have time to 
bind with light quarks to form
a bound state\cite{top}.
Therefore, the final frontier for the study of heavy-quark-antiquark bound 
states may be the $ B_c $ and $ \bar{B}_c $, which carry flavors explicitly 
and are the ground states of bound $\bar{b}c$ and $ c\bar{b}$ systems. The 
properties of these mesons and the possibility to find them at existing 
accelerators have been discussed in Refs.\cite{function}\cite{bc1}. 
The top rare
decay to $ \bar{B}_c^* $ is also an important 
channel to study the still undiscovered
heavy quark-antiquark bound state at LHC in the near future, 
though it is not practical at present
with limited top quark events.

In this paper we present the full leading order calculation of top quark 
decays into ${\bar B^*_c}$ and $\Upsilon$ in strong coupling constant 
$\alpha_s$ as well as in $v^2$, where $ v$ is the typical relative velocity of the 
heavy quarks inside the bound state. The reliance of the branching ratios
of heavy quarkonium and heavy  meson production in top decay on the mass of top 
quark is also studied.

\begin{center}\section {Formalism}\end{center}

The amplitudes for 
$t\rightarrow W^+c\bar B^*_c $ and $t\rightarrow W^+b\Upsilon$
involve two Feynman diagrams respectively as shown in Fig.1. They can be 
writen down using standard Feynman rules for bound state productions and 
decays\cite{kuhn}\cite{app2} or from the description of Bethe-Salpeter equation 
for bound states. In the following we 
will start with discussion from the latter concept.

\begin{picture}(35000,15000)
\drawline\fermion[\E\REG](1500,3000)[5000]
\drawarrow[\E\ATTIP](\pmidx,\pmidy)
\global\advance\pfrontx by -1000
\put(\pfrontx,\pfronty){$T$} 
\global\advance\pmidy by -1000
\put(\pmidx,\pmidy){$ t $}
\global\advance\pmidy by -4500
\put(\pmidx,\pmidy){$(a)$}
\drawline\fermion[\NE\REG](\pbackx,\pbacky)[2000]
\global\advance\pbackx by -500
\global\advance\pbacky by 1000 
\put(\pbackx,\pbacky){$b$}
\drawline\photon[\SE\REG](\pfrontx,\pfronty)[8]
\global\advance\pmidx by -1500
\global\advance\pmidy by -1200
\put(\pmidx,\pmidy){$W^+$}
\global\advance\pbackx by 500
\global\advance\pbacky by -500
\put(\pbackx,\pbacky){$k$}
\drawline\fermion[\NE\REG](\fermionbackx,\fermionbacky)[6000]
\drawarrow[\NE\ATTIP](\pmidx,\pmidy)
\global\advance\pbackx by -1000
\global\advance\pbacky by 1000
\put(\pbackx,\pbacky){$\bar{B}^*_c (\Upsilon)$}
\drawline\gluon[\E\FLIPPEDCENTRAL](\pfrontx,\pfronty)[2]
\drawline\fermion[\E\REG](\pbackx,\pbacky)[4000]
\drawarrow[\E\ATTIP](\pmidx,\pmidy)
\global\advance\pmidy by -1000
\put(\pmidx,\pmidy){$c(b)$}
\global\advance\pbackx by 500
\put(\pbackx,\pbacky){$p^{\prime}$}
\drawline\fermion[\NE\REG](\pfrontx,\pfronty)[4000]
\drawarrow[\SW\ATTIP](\pmidx,\pmidy)
\drawline\fermion[\NW\REG](\pbackx,\pbacky)[2050]
\global\advance\pbackx by -25
\global\advance\pbacky by -25
\drawline\fermion[\SE\REG](\pbackx,\pbacky)[2050]
\global\advance\pbackx by -25
\global\advance\pbacky by -25
\drawline\fermion[\NW\REG](\pbackx,\pbacky)[2050]
\global\advance\pbackx by -25
\global\advance\pbacky by -25
\drawline\fermion[\SE\REG](\pbackx,\pbacky)[2050]
\global\advance\pbackx by -25
\global\advance\pbacky by -25
\drawline\fermion[\NW\REG](\pbackx,\pbacky)[2050]
\global\advance\pbackx by -25
\global\advance\pbacky by -25
\drawline\fermion[\SE\REG](\pbackx,\pbacky)[2050]
\global\advance\pbackx by -25
\global\advance\pbacky by -25
\drawline\fermion[\NW\REG](\pbackx,\pbacky)[2050]
\global\advance\pbackx by -25
\global\advance\pbacky by -25
\drawline\fermion[\SE\REG](\pbackx,\pbacky)[2050]
\global\advance\pbackx by -25
\global\advance\pbacky by -25
\drawline\fermion[\NW\REG](\pbackx,\pbacky)[2050]
\global\advance\pbackx by -25
\global\advance\pbacky by -25
\drawline\fermion[\SE\REG](\pbackx,\pbacky)[2050]
\THICKLINES
\drawline\fermion[\NE\REG](\pmidx,\pmidy)[1500] 
\drawarrow[\NE\ATBASE](\pbackx,\pbacky)
\global\advance\pbackx by 500
\global\advance\pbacky by -500
\put(\pbackx,\pbacky){$P$}

\THINLINES
\drawline\fermion[\E\REG](25000,3000)[4500]
\global\advance\pfrontx by -1000
\put(\pfrontx,\pfronty){$T$} 
\drawarrow[\E\ATTIP](\pmidx,\pmidy)
\global\advance\pmidy by -1000
\put(\pmidx,\pmidy){$t$}
\global\advance\pmidy by -4700
\put(\pmidx,\pmidy){$(b)$}
\drawline\gluon[\NE\REG](\pbackx,\pbacky)[2]
\drawline\fermion[\NE\REG](\pbackx,\pbacky)[3000]
\drawarrow[\SW\ATTIP](\pmidx,\pmidy)
\global\advance\pbackx by -2000
\global\advance\pbacky by 1000
\put(\pbackx,\pbacky){$\bar{B}^*_c (\Upsilon)$}
\drawline\fermion[\NW\REG](\pfrontx,\pfronty)[3500]
\drawarrow[\NW\ATTIP](\pmidx,\pmidy)
\global\advance\pmidy by -500
\global\advance\pmidx by -2000
\put(\pmidx,\pmidy){$c(b)$}
\global\advance\pbacky by 750
\global\advance\pbackx by -500
\put(\pbackx,\pbacky){$p^{\prime}$}
\drawline\fermion[\E\REG](\gluonfrontx,\gluonfronty)[2000]
\drawline\photon[\SE\REG](\pbackx,\pbacky)[7]
\global\advance\pmidx by -1500
\global\advance\pmidy by -1200
\put(\pmidx,\pmidy){$W^+$}
\global\advance\pbackx by 500
\global\advance\pbacky by -500
\put(\pbackx,\pbacky){$k$}
\drawline\fermion[\NE\REG](\pfrontx,\pfronty)[5800]
\drawarrow[\NE\ATTIP](\pmidx,\pmidy)
\global\advance\pmidx by 500
\global\advance\pmidy by -800
\put(\pmidx,\pmidy){$b$}
\drawline\fermion[\NW\REG](\pbackx,\pbacky)[1430]
\global\advance\pbackx by -25
\global\advance\pbacky by -25
\drawline\fermion[\SE\REG](\pbackx,\pbacky)[1430]
\global\advance\pbackx by -25
\global\advance\pbacky by -25
\drawline\fermion[\NW\REG](\pbackx,\pbacky)[1430]
\global\advance\pbackx by -25
\global\advance\pbacky by -25
\drawline\fermion[\SE\REG](\pbackx,\pbacky)[1430]
\global\advance\pbackx by -25
\global\advance\pbacky by -25
\drawline\fermion[\NW\REG](\pbackx,\pbacky)[1430]
\global\advance\pbackx by -25
\global\advance\pbacky by -25
\drawline\fermion[\SE\REG](\pbackx,\pbacky)[1430]
\global\advance\pbackx by -25
\global\advance\pbacky by -25
\drawline\fermion[\NW\REG](\pbackx,\pbacky)[1430]
\global\advance\pbackx by -25
\global\advance\pbacky by -25
\drawline\fermion[\SE\REG](\pbackx,\pbacky)[1430]
\global\advance\pbackx by -25
\global\advance\pbacky by -25
\drawline\fermion[\NW\REG](\pbackx,\pbacky)[1430]
\THICKLINES
\drawline\fermion[\NE\REG](\pmidx,\pmidy)[1500] 
\drawarrow[\NE\ATBASE](\pbackx,\pbacky)
\global\advance\pbackx by 500
\global\advance\pbacky by -500
\put(\pbackx,\pbacky){$P$}
\end{picture}
\vskip 1cm
\begin{center}
{\small Fig.1}
\end{center}
\vskip 1.5cm

Based on the Mandelstam formalism \cite{mand} the amplitude of top decays 
corresponding to diagrams Fig.1 is written as
\begin{equation}
 \label{1}
 M=\frac {4 g g^2_s}{3 \sqrt 6} {\bar u}(q_2,{\bar s})\int \frac{d^4 q}
 {(2\pi)^4}\left\{ {\cal A}_1 +{ \cal A}_2 \right\} v(q_1,s).
\end{equation}
Here $ s,\bar{s}$ are the spin projections of the quark and the antiquark; 
$ q_1$
and $ q_2 $ are their momenta; $g_s$ is the coupling constant of QCD; g is 
the weak
coupling constant in WS model.

The general form of the partial width for the top decays under consideraton 
is 
\begin{equation}
\label{gamma}
\Gamma(t\rightarrow threebody)=\frac{1}{256 \pi^3 m^3_t}\int_{s_2^-}^
{s_2^+}ds_2 \int_{s_1^-}^{s_1^+}ds_1 \overline{\sum}| M^2|.
\end{equation}
In the following we will discuss the two channels of top quark decaying into vector 
mesons $\bar B_c^*$ and $\Upsilon$ which might be in the first place being 
measuerd in the future experiment on top rare decay.

\begin{center}
 \bf{A.$~~t\rightarrow \bar B_c^* +W^+ +c$}
\end{center}

At leading order in $\alpha_s$, the corresponding ${\cal A}_1, {\cal A}_2$ in Eqn.1 for
$t\rightarrow \bar B_c^* +W^+ +c $ are

\begin{equation}
\label{a1}
{\cal A}_1=\gamma_{\mu} \frac{\chi(q)}{(p'+ p_2)^2}\gamma^{\mu} \frac{\not\!{p}+ \not\!{p'}
+ m_b}{(p'+ p_2)^2 - m_b^2} {\not\!\epsilon}_{w} P_{L} ,
\end{equation}

\begin{equation}
\label{a2}
{\cal A}_2=\gamma_{\mu} \frac{\chi(q)}{(p'+ p_2)^2} {\not\!\epsilon}_w P_{L}
\frac{\not\!{p}_1 + \not\!{k} + m_b}{(p_1+k)^2 - m_t^2} \gamma^{\mu}.
\end{equation}
  Where the $ P_L=\frac{1}{2}(1-\gamma_5)$; $\chi(q)$ is the Bethe-Salpeter(BS) 
wave function of the$ \bar B_c^*$ meson with inner relative momentum q between 
the heavy quarks; $\epsilon_w$ is the polarization vector of $W^+$ boson. 
In Eqn.(\ref{a1})(\ref{a2}), $ p , k , p', p_1$ and $p_2 $ are the momenta 
of $ \bar B_c^* $ meson, $ W^+$ boson, c
quark, b quark, and $\bar{c}$ quark, respectively. For $\bar{B}_c^*$ meson, the 
two quarks have different masses inside the bound state. Their momenta $ p_1$ and  
$p_2$ have the relation

\begin{equation}
p_1=\eta_1 p +q~,~~~p_2=\eta_2 p-q,
\end{equation}
where $ \eta_1= \frac{m_b}{(m_b+m_c)}$ and $\eta_2=\frac{m_c}{(m_b+m_c)}$. 
Under 
the instantaneous approximation\cite{salpeter} with the negative energy projectors
being neglected, the BS wave function $\chi(q)$ of the 
bound state may be expressed as 

\begin{equation}
\chi(q)=\frac{i}{2 \pi}\frac{M-E_1-E_2}{(P_{10} -E_1)(P_{20}-E_2)}\Phi(\vec{q}).
\end{equation}
Here M is the mass of bound state; $ P_{10}$ and $P_{20}$ are the time 
components of quark and antiquark momenta 
inside the meson, and $ E_1, E_2$ are their
energies. 
From the standard BS wave functions in the  
approximation that the negative energy projectors are omitted, 
the vector meson wave function can be projected out as :

\begin{equation}
 \Phi(\vec{q})=\frac{1}{M}\sum_{sm}\langle JM|1sLm\rangle \Lambda^1_+ 
 (\vec{q})\gamma_0 
  \not\!{\epsilon}(M+\not\!{P})\gamma_0\Lambda^2_- (-\vec{q})\psi_{Lm}
         (\vec{q}),
\end{equation}
  where $\epsilon$ is the polarization vector associated with the spin triplet 
states. $\Lambda^1_+(\vec{q})$ and $\Lambda^2_-(-\vec{q})$ are positive energy 
projection operators of both quark and antiquark .
\begin{equation}
  \Lambda^1_+(\vec{q})=\frac{E+\gamma_0 \vec{\gamma}\cdot\vec{q}+m_c\gamma_0}{2E},~~~~
  \Lambda^2_-(-\vec{q})=\frac{E+\gamma_0 \vec{\gamma}\cdot\vec{q}-m_c\gamma_0}{2E}.
\end{equation}

After taking the nonrelativistic approximation the bound state wave function 
may be reduced to a simple form. As in Refs.\cite{app1}\cite{app2}, 
we also ignore the 
dependence of $ {\cal A}_1$ and ${\cal A}_2$ on the relative momentum $q$. It may be 
considered as the lowest order approximation. After 
considering the normalization of $ \Phi(\vec{q})$, we can get  

\begin{equation}
\int \chi(q)\frac{d^4 q}{(2 \pi)^4}=\frac{1}{4 \sqrt{M_{\bar{B}_c^*} 
\pi}}\not\!{\epsilon}
(\not\!{P} + M_{\bar{B}_c^*}) R_{\bar{B}_c^*}(0). 
\end{equation}
Here $ R_{\bar{B}_c^*}(0)$ is the radiual wave function at origin. Hence the matrix element 
squared can be carried out directly, 

\begin{equation}
\overline{\sum}\left|M\right|^2=\overline{\sum}(\left|M_1\right|^2 + \left|M_2\right|^2 + 
2 Re M_1 M_2^*).
\end{equation}
The square of the full amplitude is complicated and lengthy, it can be found in
Appendix~A. The kinematical variables entering Eq.(\ref{gamma}) are defined as 

\begin{equation}
s_1=(p'+p)^2~,~~~s_2=(p+k)^2,
\end{equation}
in which the phase space boundaries are readily found to be 
\begin{equation}
s^{\pm}_1=m_c^2+M_{\bar B_c^*}^2-\frac{1}{2 s_2}[(s_2-m_t^2+m_c^2)(s_2+
M_{\bar{B}_c^*}^2-m_w^2)\mp\lambda^{\frac{1}{2}}(s_2,m_t^2,m_c^2)
\lambda^{\frac{1}{2}}(s_2,M_{\bar B_c^*}^2,m_w^2)],            
\end{equation}
and

\begin{equation}
s_2^-=(M_{\bar B_c^*}+m_w)^2,
\end{equation}
\begin{equation}
s_2^+=(m_t-m_c)^2,
\end{equation}
where 
\begin{equation}
\lambda(x,y,z)=(x-y-z)^2-4yz.
\end{equation}

\begin{center}
{\bf$B.~~~t\rightarrow\Upsilon + W^+ + b$}
\end{center}

The ${\cal A}_1$ and ${\cal A}_2$ in Eq.(\ref{1}) 
for $t\rightarrow W^+b\Upsilon$ are the 
same form as Eqs.(\ref{a1})(\ref{a2}) 
except $ \chi(q)$ representing bottomonium BS
wave function  and $c(\bar{c})$ replaced by $ b(\bar{b})$ quark.
Under the same argument as in A, we have 

\begin{equation}
\int \chi(q)\frac{d^4 q}{(2 \pi)^4}=\frac{1}{4 \sqrt{M_{\Upsilon} \pi}}\not\!{\epsilon}
(\not\!{P} + M_{\Upsilon}) R_{\Upsilon}(0). 
\end{equation}
The matrix squared is given in Appendix B.
The phase-space bounderies to be used in  the numerical evaluation of 
Eq.(\ref{gamma}) in this channel 
are given by

\begin{equation}
s^{\pm}_1=m_b^2+m_{\Upsilon}^2-\frac{1}{2 s_2}[(s_2-m_t^2+m_b^2)(s_2+
m_{\Upsilon}^2-m_w^2)\mp\lambda^{\frac{1}{2}}(s_2,m_t^2,m_b^2)
\lambda^{\frac{1}{2}}(s_2,m_{\Upsilon}^2,m_w^2)],            
\end{equation}
and

\begin{equation}
s_2^-=(m_{\Upsilon}+m_w)^2,
\end{equation}
\begin{equation}
s_2^+=(m_t-m_b)^2.
\end{equation}

\begin{center}\section{Numerical Calculation and Results}\end{center}

We can now make use of this formalism to evaluate the decay rates of top quark
to $\Upsilon$ and $\bar{B}_c^*$ mesons. 
In numerical calculations, we take the parameters 
as follows:
\begin{center}
$m_b=4.9 GeV,~~m_c=1.5 GeV,~~m_t=176 GeV,~~m_w=80.22 GeV.$\\
$\alpha_s=0.26~~~for~ t\rightarrow W^+ c \bar{B}_c^*,$\\
\vskip -1.2cm
\begin{equation}
\alpha_s=0.19~~~for~ t\rightarrow w^+ b \Upsilon.
\end{equation}
\end{center}
Under the non-relativistic approximation, 
\begin{equation}
M_{\Upsilon}=2 m_b,~~M_{\bar{B}_c^*}=m_b+m_c .
\end{equation}
The radial wave functions at the origin for $ \bar{B}_c^*$  and $\Upsilon$ 
can be 
estimated through potential models\cite{quigg} and electric decay rate. 
As in Ref.\cite{bc2} we take
\begin{center}
$\left|R_{\bar{B}_c^*}(0)\right|^2=(1.18 GeV)^3,$\\
\vskip -1.2cm
\begin{equation}
\left|R_{\Upsilon}(0)\right|^2=(1.8 GeV)^3.
\end{equation}
\end{center}
Finally, we obtain the partial width of the decay $ t\rightarrow \bar{B}_c^* +
W^+ + c:$
\begin{equation}
\Gamma(t\rightarrow \bar{B}_c^* W^+ c)=0.64 MeV,
\end{equation}
giving
\begin{equation}
 \frac{\Gamma(t\rightarrow \bar{B}_c^* W^+ c)}{\Gamma 
 (t\rightarrow W^+ b)}=4.12\times 10^{-4}.
\end{equation}
The partial width of the decay $ t\rightarrow \Upsilon+ W^+ + b$ is
\begin{equation}
\Gamma(t\rightarrow \Upsilon + W^+ + b)=1.54\times 10^{-2} MeV,
\end{equation}
giving
\begin{equation}
 \frac{\Gamma(t\rightarrow \Upsilon W^+ b)}{\Gamma 
 (t\rightarrow W^+ b)}=0.98\times 10^{-5}.
\end{equation}

The complete leading order calculation of the 
decay rate for $ t\rightarrow W^+ b
\Upsilon$ has already been calculated\cite{app1} for  
top quark with a mass of $100 $
GeV and slightly different values for other parameters, 
giving the branching fraction $4\times 10^{-7}$. In our calculation , we 
find the branching  ratio is two order of magnitude greater when  the 
top quark mass is set to $176$ GeV. 
This indicates that the $\Upsilon$ and $\bar{B}_c^*$ meson productions in top decay is
really a quark fragmentation process if the top quark is heavy enough.
Using the fragmentation function obtained 
in Ref.\cite{bc2}, 
one can readily obtain the branching ratios for~$t\rightarrow 
\bar{B}_c^*W^+c$ and $t\rightarrow \Upsilon W^+ b$. However, in how much 
degree the univeresal fragmentation functions suit  the top rare decays is 
not yet very clear. To answer this question, we give out the 
relationships between the top quark 
mass and the branching ratios in Fig.2 and Fig.3
respectively for $t\rightarrow \Upsilon W^+ b$ and $t\rightarrow 
\bar{B}_c^*W^+c$.
From Fig.2 and Fig.3  we can see that  the branching ratios increase rapidly
as the top quark mass increases when $m_t \leq 120$ GeV, and the heavier
the top quark is, the less the branching ratios change.
When top quark mass is heavier than $200$ GeV, the  
branching ratios seem to be saturated and 
no longer depend on the top quark mass.
That means the univeral 
fragmentation functions works well.
As for the numerical results, with the same parameters as used in 
Refs.\cite{bc2} at $m_t = 176$ GeV, our complete leading order 
calculations give $R(t\rightarrow \Upsilon W^+ b)=0.98\times 10^{-5}$, which
is about two times smaller than in Ref.\cite{bc2} using the universal
fragmentation function and $R(t\rightarrow \bar{B}_c^* W^+ c)=4.12 \times 
10^{-4}$, which is almost as large as Ref.\cite{bc2} using the universal 
fragmentation function.

In Fig.4 and Fig.5 we also show the relations of these branching ratios to
the strong coupling constant $\alpha_s$.


\begin{center}\section{Conclusion}\end{center}

We have presented two dominant three-body decays of top quark to 
ground state heavy mesons.
The decay widths and branching ratios  are 
$0.64 MeV $ and $4.12\times 10^{-4}  $ for $ \bar{B}_c^* $ productionin ;   
$1.54\times 10^{-2} MeV  $ and $0.98\times 10^{-5}  $ 
for $ \Upsilon $ production . 
As about $10^8~ t\bar{t}$ pairs per year will be produced at the 
planned LHC, this means it is possible to accumulate about $10^4$ 
$\bar{B}_c^* $ 
events
and $10^3$ $\Upsilon$ events a year. This will make 
it possible to discover the $\bar{B}_c^*$ meson
and to further study its properties experimentally. At 
NLC (Next
Linear Collider), rare decays of top quark could be searched for down to a 
very small branching ratos, and a systematic and general search for top rare 
decay modes may be possible. Then the two top decay modes discussed in this 
paper might correspond to a detectable level at
NLC. On the other hand the results obtained in this paper will also be a 
helpful reference to other production mechamisms concerning the top quark 
decay.
\vskip 1cm
\begin{center}
\bf\large\bf{Acknowlegement}
\end{center}

This work was supported in part by the National Natural Science Foundation
of China, the State Education Commission of China and the State Commission
of Science and Technology of China.
\newpage
\begin{center}\bf{APPENDIX A} 
\end{center}

Following we give the expressions of squared matrix 
for $t\rightarrow W^+c\bar{B}_c^* $.
The vertex coefficents are not taken into consideration. In this appendix we 
take
$ {\widetilde M}_{\bar{B}_c^*} = \frac{1}{2} M_{\bar{B}_c^*} $ just for 
calculation convenience.
\begin{eqnarray*}
\overline{\sum}|M_{1}|^2 
    & = &
       \frac{1}{(4 m_{b}^2 {\widetilde M}_{\bar{B}_c^*}^2 m_{w}^2 + 8 m_{b} {\widetilde M}_{\bar{B}_c^*}^2 
     m_{c} m_{w}^2 + 4 {\widetilde M}_{\bar{B}_c^*}^2 m_{c}^2 m_{w}^2)} \\
    & \times & 
    \frac{1}{(\eta_1 s_2- 4 \eta_2 {\widetilde M}_{\bar{B}_c^*}^2 + \eta_2 m_{w}^2 + 
    4 \eta_2^2 {\widetilde M}_{\bar{B}_c^*}^2-m_{t}^2)^2} \frac{1}{4 \eta_2^2 
    (s_1-m_{b}^2)^2} \\
    & \times & 
     \big\{ 64 m_{b}^2 {\widetilde M}_{\bar{B}_c^*}^2 (k\cdot{p})^2 (p\cdot{p'}) (t\cdot{p})  - 
    64 m_{b} {\widetilde M}_{\bar{B}_c^*}^2 m_{t}^2 (m_b + m_c) 
    (k\cdot{p})^2 (p\cdot{p'}) \\
    & + &
       128 m_{b}^2 {\widetilde M}_{\bar{B}_c^*}^3 m_{c} (k\cdot{p})^2 
       (t\cdot{p}) 
     - 320 m_{b} {\widetilde M}_{\bar{B}_c^*}^3 m_{c} m_{t}^2 
    (m_b + m_c) (k\cdot{p}) ^2  \\
    & + &
    32 (4 m_b^2 {\widetilde M}_{\bar{B}_c^*}^2 + m_b^2 m_t^2 + 4 m_b^2 m_w^2
    + 2 m_b m_c m_t^2 \\
    & + &
    8 m_b m_c m_w^2 + m_c^2 m_t^2 + 4 m_c^2 m_w^2)
    {\widetilde M}_{\bar{B}_c^*}^2 (k\cdot{p})^2 (t\cdot{p'})  \\
    & - &
     64 {\widetilde M}_{\bar{B}_c^*}^2 ( 2 m_{b}^2 
     {\widetilde M}_{\bar{B}_c^*}^2 - m_b^2 m_t^2 - m_b^2 m_w^2 - 
     2 m_b m_c m_t^2 \\
     & - &
     2 m_b m_c m_w^2 - m_c^2 m_t^2 - m_c^2 m_w^2) 
     (k\cdot{p})  (k\cdot{p'})  (t\cdot{p})  \\
    & + &
    128  m_{b}^2 {\widetilde M}_{\bar{B}_c^*}^4 
    ( k\cdot{p})  (k\cdot{t})  (p\cdot{p'}) 
     + 64  m_{b} {\widetilde M}_{\bar{B}_c^*}^2 m_{w}^2 (m_b + m_c) 
    (k\cdot{p})  (p\cdot{p'})  (t\cdot{p}) \\
    & + &
    64 {\widetilde M}_{\bar{B}_c^*}^3 (- 4 m_{b}^2 m_{c} 
    {\widetilde M}_{\bar{B}_c^*}^2 + 
    m_{t}^2 m_b^2 m_c + 4 m_{w}^2 m_b^2 m_c + 
    2  m_{t}^2 m_c^2 m_b \\
    & + &
    8 m_{w}^2 m_b m_c^2 + m_c^3 m_t^2 + 4 m_c^3 m_w^2)
    (k\cdot{p})  (k\cdot{t}) \\
    & - &
    160 {\widetilde M}_{\bar{B}_c^*}^2 m_{t}^2 m_{w}^2 (m_b + m_c)^2 
    (k\cdot{p})  (p\cdot{p'})  \\
    & + & 
    192  m_{b} m_c {\widetilde M}_{\bar{B}_c^*}^3 m_{w}^2 (m_b + m_c)
    (k\cdot{p})  (t\cdot{p}) \\
    & + & 
    704  m_{b} {\widetilde M}_{\bar{B}_c^*}^4 m_{w}^2 (m_b + m_c)
    (k\cdot{p})  (t\cdot{p'}) \\
    & - & 
    672 m_c {\widetilde M}_{\bar{B}_c^*}^3  m_{t}^2 m_{w}^2 (m_c + m_b)^2
    (k\cdot{p}) \\
    & - &
    64 {\widetilde M}_{\bar{B}_c^*}^4 (4 m_b^2 {\widetilde M}_{\bar{B}_c^*}^2
    + m_b^2 m_t^2 - 2 m_b^2 m_w^2 + 2 m_b m_c m_t^2 
    - 4 m_b m_c m_t^2 \\
    & + & 
    m_c^2 m_t^2 - 2 m_c^2 m_w^2) (k\cdot{p'})  (k\cdot{t}) 
     + 192  m_{b} {\widetilde M}_{\bar{B}_c^*}^4 m_{w}^2 (m_b + m_c) 
    (k\cdot{p'})  (t\cdot{p}) \\
    &  - &
    128 {\widetilde M}_{\bar{B}_c^*}^4 m_{t}^2 m_{w}^2 (m_b + m_c)^2 
    (k\cdot{p'}) 
    + 768 m_{b} m_{c} {\widetilde M}_{\bar{B}_c^*}^5 m_{w}^2 (m_b + m_c) 
    (k\cdot{t}) \\
    & + &
    192  m_{b} {\widetilde M}_{\bar{B}_c^*}^4 m_{w}^2 (m_b + m_c) 
    (k\cdot{t}) (p\cdot{p'}) \\
    & + &
    16 {\widetilde M}_{\bar{B}_c^*}^2 m_{w}^2 
    (8 {\widetilde M}_{\bar{B}_c^*}^2 
    m_{b}^2 + 2 m_b^2 m_t^2 - m_b^2 m_w^2 \\
    & + & 
    4 m_b m_c m_t^2 
    - 2 m_b m_c m_w^2 
    + 2 m_c^2 m_t^2 - m_c^2 m_w^2) (p\cdot{p'}) (t\cdot{p})  \\
    & - &
    512 m_{b} {\widetilde M}_{\bar{B}_c^*}^4 m_{t}^2 m_{w}^2 (m_b + m_c) 
    (p\cdot{p'})\\
    & + & 
    64 {\widetilde M}_{\bar{B}_c^*}^3 m_{c} m_{w}^2 (
    8 m_{b}^2 {\widetilde M}_{\bar{B}_c^*}^2 + 2 m_{b}^2 m_{t}^2
    - m_{b}^2 m_{w}^2 \\
    & + & 
    4 m_b m_c m_t^2 - 2 m_b m_c m_w^2 + 2 m_c^2 m_t^2
    - m_c^2 m_w^2) (t\cdot{p}) \\
    & + & 
    32 {\widetilde M}_{\bar{B}_c^*}^4 m_{w}^2 ( 28 m_{b}^2 
    {\widetilde M}_{\bar{B}_c^*}^2 + 7 m_b^2 m_t^2 \\
    & - & 5 m_b^2 m_w^2 + 
    17 m_b m_c m_t^2 - 10 m_b m_c m_w^2 + 7 m_c^2 m_t^2 - 5 m_c^2 m_w^2)
    (t\cdot{p'}) \\
    & - &
    1408 m_{b} m_c {\widetilde M}_{\bar{B}_c^*}^5 m_{t}^2 m_{w}^2 (m_b + 
    m_c)\big\} \\
\end{eqnarray*}   

\begin{eqnarray*}
\overline{\sum}|M_{2}|^2 
    & = &
    \frac{1}{4 \eta_2^2 (s_1-m_{b}^2)^2} 
    \frac{1}{(s_1-m_b^2)^2}\\ 
    & \times & 
    \big\{(8 {\widetilde M}_{\bar{B}_c^*}^2 - m_c^2 + m_b^2) (p\cdot{p'}) (t\cdot{p})\\
    & + &
     2 {\widetilde M}_{\bar{B}_c^*}^2 (-4 {\widetilde M}_{\bar{B}_c^*}^2 - 8 {\widetilde M}_{\bar{B}_c^*} 
     m_b + 8 {\widetilde M}_{\bar{B}_c^*} m_c + m_c^2 +  m_b^2-
     6 m_b m_c) (t\cdot{p'})\\
    & + &
    4 {\widetilde M}_{\bar{B}_c^*}^2 (m_c^2 - 3  m_b m_c) (t\cdot{p}) + 
    8 {\widetilde M}_{\bar{B}_c^*}^2 (p\cdot{p'}) (t\cdot{p'})+2 (p\cdot{p'})^2 (t\cdot{p'})\\
    & + &
     \frac{4 {\widetilde M}_{\bar{B}_c^*}^2}{m_w^2} (-4 {\widetilde M}_{\bar{B}_c^*}^2 + 
      m_c^2+ m_b^2 - 6  m_b m_c) (k\cdot{t}) (k\cdot{p'}) \\
    & + & 
     \frac{16 {\widetilde M}_{\bar{B}_c^*}^2}{m_w^2} 
     (k\cdot{t}) (k\cdot{p'}) (p\cdot{p'}) 
     + 
     \frac{2}{m_w^2} 
     (8 {\widetilde M}_{\bar{B}_c^*}^2+ m_b^2 - m_c^2) (k\cdot{t}) (k\cdot{p}) (p\cdot{p'})\\
    & + & 
     \frac{8 {\widetilde M}_{\bar{B}_c^*}^2}{m_w^2} 
     ( m_c^2 - 3 m_b m_c) (k\cdot{t}) (k\cdot{p}) +
    \frac{4}{m_w^2} (k\cdot{t}) (k\cdot{p'}) (p\cdot{p'})^2)\big\}\\
\end{eqnarray*}   

\begin{eqnarray*}
\overline{\sum}2 Re (M_{1}M_{2}^*) 
    & = &
       \frac{1}{ {\widetilde M}_{\bar{B}_c^*} m_{w}^2 (m_{b}  +  m_{c}) } 
       \frac{1}{(s_1-m_b^2)}\frac{1}{4 \eta_2^2 (s1-m_{b}^2)^2} \\
    & \times & 
    \frac{1}{(\eta_1 s_2 - 4 \eta_2 {\widetilde M}_{\bar{B}_c^*}^2 + \eta_2 m_{w}^2 + 
    4 \eta_2^2 {\widetilde M}_{\bar{B}_c^*}^2-m_{t}^2)} \\
    & \times & 
    \big\{
   8 m_b {\widetilde M}_{\bar{B}_c^*}  (k\cdot{p})^2 
   (p\cdot{p'}) (t\cdot{p'}) 
    - 8 m_b {\widetilde M}_{\bar{B}_c^*}^2 (m_b + 2 {\widetilde M}_{\bar{B}_c^*} 
    - m_c) (k\cdot{p})^2 (t\cdot{p'})\\
   & + &
   4 {\widetilde M}_{\bar{B}_c^*} m_c m_t^2 (m_b^2 + 4 m_b 
   {\widetilde M}_{\bar{B}_c^*} + 4 m_c {\widetilde M}_{\bar{B}_c^*} -m_c^2) 
   (k\cdot{p})^2 \\
   & - &
   8 m_b {\widetilde M}_{\bar{B}_c^*} (k\cdot{p}) 
   (k\cdot{p'}) (p\cdot{p'}) (t\cdot{p}) 
    - 32 (k\cdot{p}) (k\cdot{p'}) (t\cdot{p'}) m_b 
   {\widetilde M}_{\bar{B}_c^*}^3 \\
   & + &
   8 {\widetilde M}_{\bar{B}_c^*} m_t^2 (m_b + m_c) 
   (k\cdot{p}) (k\cdot{p'}) (p\cdot{p'})  \\
   & + &
   8  m_b {\widetilde M}_{\bar{B}_c^*}^2 (m_b + 2  
   {\widetilde M}_{\bar{B}_c^*} - m_c) (k\cdot{p}) (k\cdot{p'}) (t\cdot{p}) \\
   & - &
   16 {\widetilde M}_{\bar{B}_c^*}^2 m_t^2 (m_b^2 + m_b 
   {\widetilde M}_{\bar{B}_c^*} + m_c {\widetilde M}_{\bar{B}_c^*} - m_c^2) 
   (k\cdot{p}) (k\cdot{p'})  \\
   & - &
   8 m_b {\widetilde M}_{\bar{B}_c^*} (k\cdot{p}) (k\cdot{t}) (p\cdot{p'})^2  
    + 32 m_b m_c {\widetilde M}_{\bar{B}_c^*}^3 (m_b + 
   {\widetilde M}_{\bar{B}_c^*}) (k\cdot{p}) (k\cdot{t}) \\
   & - &
   8 m_b {\widetilde M}_{\bar{B}_c^*}^2 (m_b + 6 {\widetilde M}_{\bar{B}_c^*} 
   + m_c) (k\cdot{p}) (k\cdot{t}) (p\cdot{p'}) \\
   & + &
   4 {\widetilde M}_{\bar{B}_c^*} m_w^2 ( m_b + m_c) (k\cdot{p}) (p\cdot{p'}) (t\cdot{p'})  \\
   & + &
   8 {\widetilde M}_{\bar{B}_c^*} m_c m_w^2 (m_b^2 + 4 m_b 
   {\widetilde M}_{\bar{B}_c^*} + 4 {\widetilde M}_{\bar{B}_c^*} m_c - m_c^2) 
   (k\cdot{p}) (t\cdot{p})  \\
   & - & 
   8 {\widetilde M}_{\bar{B}_c^*}^2 m_w^2 (3 m_b^2 + 5 m_b 
   {\widetilde M}_{\bar{B}_c^*} + 2 m_b m_c + 5 
   {\widetilde M}_{\bar{B}_c^*} m_c - m_c^2) (k\cdot{p}) (t\cdot{p'})  \\
   & + &
   32 m_b {\widetilde M}_{\bar{B}_c^*}^3 (k\cdot{p'})^2 (t\cdot{p})  - 
   32 {\widetilde M}_{\bar{B}_c^*}^3 m_t^2 (m_b + m_c) (k\cdot{p'})^2  \\
   & + &
   64 m_b {\widetilde M}_{\bar{B}_c^*}^4 ({\widetilde M}_{\bar{B}_c^*} + 
   m_c) (k\cdot{p'}) (k\cdot{t})  \\
   & + &
   12 {\widetilde M}_{\bar{B}_c^*} m_w^2 ( m_b + m_c) 
   (k\cdot{p'}) (p\cdot{p'}) (t\cdot{p})  \\
   & + &
   8 {\widetilde M}_{\bar{B}_c^*}^2 m_w^2 (-m_b^2 + m_b 
   {\widetilde M}_{\bar{B}_c^*} + 2 m_b m_c + m_c 
   {\widetilde M}_{\bar{B}_c^*} + 3 m_c^2) (k\cdot{p'}) (t\cdot{p})  \\
   & - &
   64 {\widetilde M}_{\bar{B}_c^*}^3 m_w^2 (m_b + m_c) 
   (k\cdot{p'}) (t\cdot{p'})  
   - 4 {\widetilde M}_{\bar{B}_c^*} m_w^2 (m_b + m_c) 
   (k\cdot{t}) (p\cdot{p'})^2 \\
   & + &
   8 {\widetilde M}_{\bar{B}_c^*}^2 m_w^2 (m_b^2 + m_b 
   {\widetilde M}_{\bar{B}_c^*} + m_c {\widetilde M}_{\bar{B}_c^*} - m_c^2)
   (k\cdot{t}) (p\cdot{p'}) \\
   & - &
   8 {\widetilde M}_{\bar{B}_c^*}^3 m_c m_w^2 (m_b^2 + 4 m_b 
   {\widetilde M}_{\bar{B}_c^*} - 2 m_b m_c + 4 
   {\widetilde M}_{\bar{B}_c^*} m_c - 3 m_c^2) (k\cdot{t})  \\
   & + &
   16 m_b {\widetilde M}_{\bar{B}_c^*} m_w^2 (p\cdot{p'})^2 (t\cdot{p}) 
    - 8 {\widetilde M}_{\bar{B}_c^*} m_t^2 m_w^2 (m_b + m_c) (p\cdot{p'})^2 \\
   & + &
   16 m_b {\widetilde M}_{\bar{B}_c^*}^2 m_w^2 (m_c + 
   2 {\widetilde M}_{\bar{B}_c^*}) (p\cdot{p'}) (t\cdot{p})  \\
   & - &
   48 m_b {\widetilde M}_{\bar{B}_c^*}^3 m_w^2 (p\cdot{p'}) (t\cdot{p'}) -
   4 {\widetilde M}_{\bar{B}_c^*}^2 m_t^2 m_w^2 (-m_b^2 + 8 m_b 
   {\widetilde M}_{\bar{B}_c^*} \\
   & + & 
   8 m_c {\widetilde M}_{\bar{B}_c^*} + m_c^2) (p\cdot{p'})  
    - 32 m_b {\widetilde M}_{\bar{B}_c^*}^4 m_w^2 (2 m_b + 
   3 {\widetilde M}_{\bar{B}_c^*}) (t\cdot{p'}) \\
   & - &
   8 m_b m_c {\widetilde M}_{\bar{B}_c^*}^3 m_w^2 (m_b - 8 
   {\widetilde M}_{\bar{B}_c^*} + m_c ) (t\cdot{p})  \\
   & + & 
   16 m_c {\widetilde M}_{\bar{B}_c^*}^3 m_t^2 m_w^2 (2 m_b^2  - 
   m_b {\widetilde M}_{\bar{B}_c^*} + 2 m_b m_c - 
   m_c {\widetilde M}_{\bar{B}_c^*})\big\}\\
\end{eqnarray*}   
In the above
\vskip -2cm
\begin{flushleft}
\begin{eqnarray*}
k\cdot p\equiv p\cdot k = \frac{1}{2}(s_2 -4\widetilde{M}_{\bar{B}_c^*}^2 -
m_w^2),~~~& &p\cdot p'= \frac{1}{2}(s_1 - m_c^2 - 4\widetilde{M}_{\bar{B}_c^*}^2)\\
p\cdot t\equiv t\cdot p = p\cdot p' + p\cdot k + 4\widetilde{M}_{\bar{B}_c^*}^2,
~~~& &k\cdot t\equiv t\cdot k = -\frac{1}{2}(s_1 - m_t^2 - m_w^2)\\k\cdot p' = k\cdot t - k\cdot p - m_w^2,~~~& &t\cdot p'= -\frac{1}{2}(s_2 - m_t^2
- m_c^2)
\end{eqnarray*}
\end{flushleft}
\vskip 1cm

\begin{center}{\bf APPENDIX B}\end{center}

In this appendix we give the expressions of squared matrix for $t\rightarrow
W^+ + b + \Upsilon $. The vertex coefficients are also neglected.

\begin{eqnarray*}
\overline{\sum}|M_{1}|^2 
    & = &\frac{1}{(s_2+m_w^2-2 (m_b^2+m_t^2))^2}\frac{1}{(s_1-2 m_b^2+m_b^2)^2}\\
    & \times &
       \big\{ \frac{4}{m_w^2} (k\cdot{p})^2 (p\cdot{p'}) (t\cdot{p})-
        \frac{8 m_t^2}{m_w^2} (k\cdot{p})^2 (p\cdot{p'})
        + \frac{8 m_b^2}{m_w^2} (k\cdot{p})^2 (t\cdot{p})\\
        & + &
        \frac{8}{m_w^2}
         (m_b^2 + m_t^2+ 4 m_w^2) (k\cdot{p})^2 (t\cdot{p'})
         -  
         \frac{40}{m_w^2} m_t^2 m_b^2 (k\cdot{p})^2\\
         & + &
        \frac{8}{m_w^2}(-m_b^2 + 2 m_t^2 + 2 m_w^2) (k\cdot{p}) (k\cdot{p'}) (t\cdot{p})
         +  
         \frac{8 m_b^2}{m_w^2} (k\cdot{p}) (t\cdot{k}) (p\cdot{p'})\\
        & + &
        \frac{16 m_b^2}{m_w^2}(-m_b^2 + m_t^2  + 
        4 m_w^2) (k\cdot{p}) (t\cdot{k})
         + 
        8 (k\cdot{p}) (p\cdot{p'}) (t\cdot{p}) \\
        & - &
        40 m_t^2 (k\cdot{p}) (p\cdot{p'})
         +  
        24 m_b^2 (k\cdot{p}) (t\cdot{p}) +
        88 m_b^2 (k\cdot{p}) (t\cdot{p'})\\
        & - &
        168 m_b^2 m_t^2 (k\cdot{p}) -
        \frac{16 m_b^2}{m_w^2} (m_b^2 +  m_t^2 - 2 m_w^2) 
        (k\cdot{p'}) (t\cdot{k}) \\
        & + &
        24 m_b^2 (k\cdot{p'}) (t\cdot{p}) 
         - 
        32 m_b^2 m_t^2 (k\cdot{p'})
         + 
        24 m_b^2 (t\cdot{k}) (p\cdot{p'})\\
        & + &
        96 m_b^4 (t\cdot{k})
         + 
        4 (2 m_b^2 + 2 m_t^2 - m_w^2) (p\cdot{p'}) (t\cdot{p}) 
         - 
        64 m_b^2 m_t^2 (p\cdot{p'})\\
        & + &
        16 m_b^2 ( 2 m_b^2 + 2 m_t^2 - m_w^2 ) (t\cdot{p})
         + 
        8 m_b^2 (7 m_b^2 + 7 m_t^2 - 5 m_w^2) (t\cdot{p'})- 
        176 m_b^4 m_t^2 \big\} \\
\end{eqnarray*}   
\begin{eqnarray*}
\overline{\sum}|M_{2}|^2 
    & = &\frac{1}{4 (s_1- m_b^2)^4}\\
    & \times &
       \big\{
           8 m_b^2 (p\cdot{p'}) (t\cdot{p})
           -16 m_b^4  (t\cdot{p'})
           -8 m_b^4 (t\cdot{p})\\
           & + &
           8 m_b^2 (p\cdot{p'}) (t\cdot{p'}) + 
           2 (p\cdot{p'})^2 (t\cdot{p'})\\
           & - &
           \frac{32 m_b^4}{m_w^2} (k\cdot{t}) (k\cdot{p'})
             - 
           \frac{16 m_b^4}{m_w^2} (k\cdot{t}) (k\cdot{p}) \\
           & + &
           \frac{16 m_b^2}{m_w^2} (k\cdot{t}) (k\cdot{p'}) (p\cdot{p'})\\
           & + &
           \frac{16 m_b^2}{m_w^2} (k\cdot{t}) (k\cdot{p}) (p\cdot{p'}) + 
           \frac{4}{m_w^2} (k\cdot{t}) (k\cdot{p'}) (p\cdot{p'})^2\big\}\\
\end{eqnarray*}   
\begin{eqnarray*}
\overline{\sum}2 Re (M_{1}M_{2}^*) 
    & = &
     \frac{1}{(s_1-m_b^2)^3}\frac{1}{(s_2 + m_w^2 - 2 (m_b^2 + m_t^2))} \\
   & \times &
     \big\{\frac{2}{m_w^2} (k\cdot{p})^2 (p\cdot{p'}) (t\cdot{p'})
      - 
    \frac{ 4 m_b^2 }{m_w^2} 
    (k\cdot{p})^2 (t\cdot{p'})
     + 
    \frac{ 8 m_t^2 m_b^2}{m_w^2} (k\cdot{p})^2 \\
     & - &
    \frac{2}{m_w^2} (k\cdot{p}) (k\cdot{p'}) (p\cdot{p'}) (t\cdot{p})+
    \frac{4 m_t^2}{m_w^2} (k\cdot{p}) (k\cdot{p'}) (p\cdot{p'}) \\
    & + & 
    \frac{ 4 m_b^2 }{m_w^2}
    (k\cdot{p}) (k\cdot{p'}) (t\cdot{p}) 
     - 
    \frac{ 8 m_b^2}{m_w^2}(k\cdot{p}) (k\cdot{p'}) (t\cdot{p'}) \\
    & - &
    \frac{8 m_t^2 m_b^2}{m_w^2}
     (k\cdot{p}) (k\cdot{p'}) 
     - 
    \frac{2}{m_w^2} (k\cdot{p}) (k\cdot{t}) (p\cdot{p'})^2 \\
    & - &
    \frac{16 m_b^2}{m_w^2}
     (k\cdot{p}) (k\cdot{t}) (p\cdot{p'}) 
     +  
    \frac{16 m_b^4}{m_w^2}
    (k\cdot{p}) (k\cdot{t}) \\
    & + &
    2 (k\cdot{p}) (p\cdot{p'}) (t\cdot{p'}) 
     + 
    16 m_b^2 (k\cdot{p}) (t\cdot{p}) 
     - 
    28  m_b^2 
    (k\cdot{p}) (t\cdot{p'}) \\
    & + &
    \frac{8 m_b^2}{m_w^2} (k\cdot{p'})^2 (t\cdot{p}) 
     - 
    \frac{16 m_t^2 m_b^2}{m_w^2} (k\cdot{p'})^2 +
    \frac{32 m_b^4 }{m_w^2} 
    (k\cdot{p'}) (k\cdot{t}) \\
    & + &
    6 (k\cdot{p'}) (p\cdot{p'}) (t\cdot{p}) +
     12 m_b^2 
    (k\cdot{p'}) (t\cdot{p}) 
     - 
    32 m_b^2 (k\cdot{p'}) (t\cdot{p'}) \\
    & - & 
    2 (k\cdot{t}) (p\cdot{p'})^2 
     + 
    4 m_b^2  
    (k\cdot{t}) (p\cdot{p'}) 
     - 
    8 m_b^4 
    (k\cdot{t}) \\
    & + &
    4 (p\cdot{p'})^2 (t\cdot{p}) - 4 m_t^2 (p\cdot{p'})^2 + 
    12 m_b^2 (p\cdot{p'}) (t\cdot{p})\\
    & - &
    12 m_b^2 (p\cdot{p'}) (t\cdot{p'}) -
    16 m_b^2  m_t^2 (p\cdot{p'}) 
     + 
    12 m_b^4 
    (t\cdot{p})\\ 
    & - &
    40 m_b^4 (t\cdot{p'}) + 
    8 m_b^4 m_t^2 \big\}\\
\end{eqnarray*}   
Here
\vskip -2cm
\begin{flushleft}
\begin{eqnarray*}
k\cdot p\equiv p\cdot k = \frac{1}{2}(s_2 -4m_b^2 -
m_w^2),~~~& &p\cdot p'= \frac{1}{2}(s_1 - 5m_b^2 )\\
p\cdot t\equiv t\cdot p = p\cdot p' + p\cdot k + 4m_b^2,
~~~& &k\cdot t\equiv t\cdot k = -\frac{1}{2}(s_1 - m_t^2 - m_w^2)\\k\cdot p' = k\cdot t - k\cdot p - m_w^2,~~~& &t\cdot p'= -\frac{1}{2}(s_2 - m_t^2
- m_b^2)
\end{eqnarray*}
\vskip 3cm
\end{flushleft}

\vfill\eject

\newpage
\centerline{\Large \bf Figure Captions}
\par
\vskip 6mm
\noindent
The Feynman diagrams for $t\rightarrow \bar{B}_c^* W^+ c$ 
and $t\rightarrow \Upsilon W^+ b$  process at leading order in $\alpha_s$.
\par
\vskip 6mm
\noindent
Fig.2 The branching ratio $R\equiv\Gamma(t\rightarrow\Upsilon+ 
W^{+}+b)/\Gamma(t\rightarrow{W^+}b)$ versus $m_t (GeV)$
\par
\vskip 6mm
\noindent
Fig.3 The branching ratio $R\equiv\Gamma(t\rightarrow{\bar{B}_c^*}+
W^{+}+c)/\Gamma(t\!\rightarrow\!{W^+}b)$ versus $m_t (GeV)$
\par
\vskip 6mm
\noindent
Fig.4 The branching ratio $R\equiv\Gamma(t\rightarrow\Upsilon+
W^{+}+b)/\Gamma(t\rightarrow{W^+}b)$ versus $\alpha_s$
\par
\vskip 6mm
\noindent
Fig.5 The branching ratio $R\equiv\Gamma(t\rightarrow{\bar{B}_c^*}+
W^{+}+c)/\Gamma(t\rightarrow{W^+}b)$ versus $\alpha_s$

\newpage
\begin{thebibliography}{99}
\bibitem{s1} W. Hollik, in Proceedings of the XVI International Symposium on
  Lepton-Photon Interactions, Connell University, Ithaca, N.Y., Aug. 10-15 1993;
  M. Swartz, in Proceedings of the XVI International Symposium on
  Lepton-Photon Interactions, Connell University, Ithaca, N.Y., Aug. 10-15 1993.
\bibitem{s2} G. Altarelli, in Proceedings of Interational University School of
  Nuclear and Partical Physics: Substructures of Matter as Revealed with  
  Electroweak Probes, Schladming, Ausria, 24 Feb - 5 Mar. 1993. 
\bibitem{s3} G. Altarelli, CERN-TH-7319/94, talk at 1st International Conference
  on Phenomenology of Unification: from Present to Future, Rome, Itali, 23-26 
  Mar 1994.
\bibitem{s4} G. L. Kane, in Proceedings of the Workshop on High Energy Phenomenology,
  Mexico City, July 1-10, 1991.
\bibitem{s5} F. Abe, \cal {et al}. (CDF Collaboration), Phys. Rev. Lett. {\bf 74},
  2626 (1995).
\bibitem{s6} S. Abachi, \cal {et al}. (D0 Collaboration), Phys. Rev. Lett. {\bf74},
  2632 (1995).
\bibitem{LHC} V. Barger and R. J. Phillips, Preprint MAD/PH/789, 1993.
\bibitem{function} E. Braaten and T.C. Yuan, Phys. Rev. Lett. {\bf 71} (1993) 1673;
 Phys. Rev. D{\bf50} (1994) 3176;
 C.-H. Chang and Y.-Q. Chen, Phys. Lett. {\bf B284}, 127 (1992); 
 Phys. Rev. {\bf D46}, 3845 (1992);
 Y.-Q. Chen, Phys. Rev. {\bf D48} (1993) 5181;
 T.C. Yuan, Phys. Rev. {\bf D50} (1994) 5664.    
\bibitem{bc2} 
 E. Braaten, K. Cheung and T.C. Yuan, Phys. Rev. {\bf D48} (1993) 4230;
 Phys. Rev. {\bf D48} (1993) R5049.
\bibitem{nrqcd} G. T. Bowdin, E. Braaten, and G. P. Lepage, Phys. Rev. D{\bf 51}, 1125 (1995).
\bibitem{oct} E. Braaten, and S. Fleming, Phys. Rev. Lett. {\bf 74} (1995) 
 3327; hep-ph/9507398 (1995).
\bibitem{kuhn} J.H. K{\rm $\ddot{u}$}hn, J. Kaplan and E.G.O. Safiani, 
  Nucl. Phys. B{\bf 157} (1979) 125. 
\bibitem{top} I.I. Bigi, Yu L. Dokshitzer, V.A. Khoze, 
 J.H. K{\rm $\ddot{u}$}hn
   and P. Zerwas, Phys. Lett. {\bf181B} (1986) 157; L.H. Orr and J.L. Rosner, 
   Phys. Lett. {\bf246B} (1990)221; {\bf248B} (1990) 474(E).
\bibitem{bc1} Y.-Q. Chen, Ph.D. thesis, Academia Sinica; C.-H Chang and Y.-Q.
   Chen, Phys. Rev. D{\bf49}, 3399(1994); E. Eichten and C. Quigg, ibid. {\bf49}
   5845(1994); V.V. Kiselev, A.K. Likhoded, and A.V. Tkabladze, ibid {\bf51},
   3613 (1995); C.-H. Chang and Y.-Q. Chen, ibid. {\bf48}, 4086 (1993);
  E. Braaten, K. Cheung and T.C. Yuan, ibid. {\bf48}, 5049 (1993);  
  K. Cheung, Phys. Rev. Lett. {\bf71}, 3413 (1993).
\bibitem{salpeter} E.E. Salpeter, Phys. Rev. {\bf87}, 328(1952).
\bibitem{mand} S. Mandelstam, Proc. R. Soc. London A{\bf223}, 248(1955).
\bibitem{quigg} E.J. Eichten and C. Quigg, preprint Fermilab-pub-95/045-T.
\bibitem{app1} V. Barger, K. Cheung, and W.-Y. Keung, Phys. Rev. D{\bf41}, 1541 (1990).
\bibitem{app2} B. Guberina, J.H. K{\rm $\ddot{u}$}hn, R.D. Peccei, and 
    R. R{\rm $\ddot{u}$}ckl, 
    Nucl. Phys. {\bf B174} (1980), 317.
\end{thebibliography}
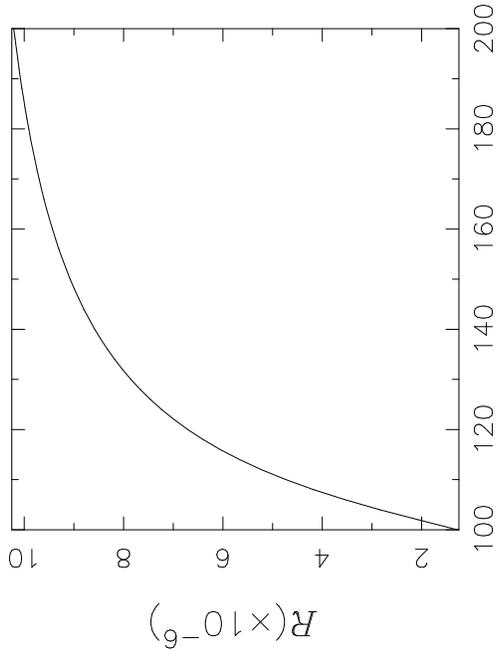
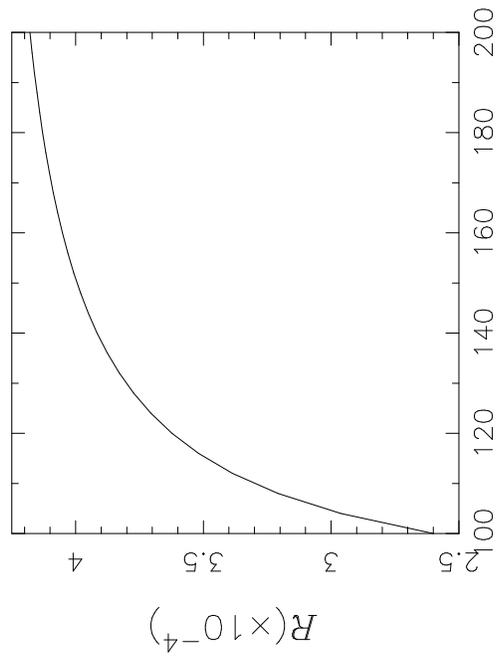
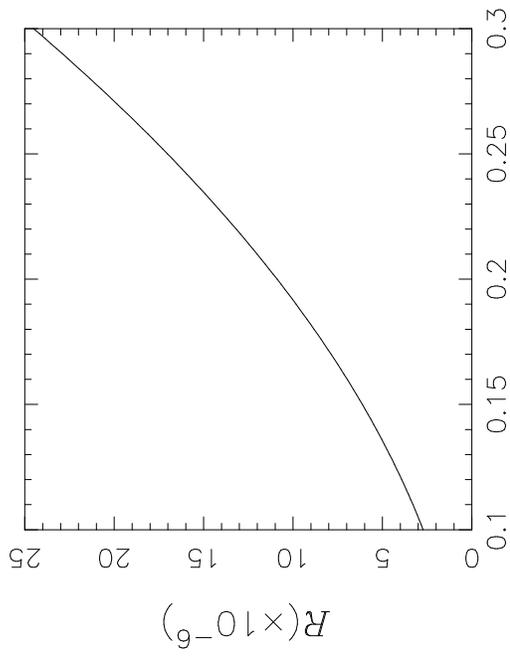
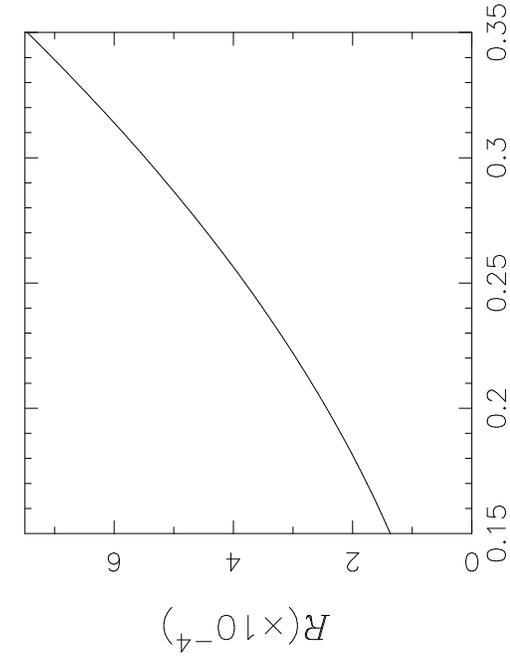
\end{document}